\documentclass[twocolumn,           % Format : preprint, twocolumn
               showpacs,            % Pacs : showpacs, noshowpacs
               nopreprintnumbers,     % Preprint: preprintnumbers,
               aps,                 % Society: ...
               prd,          	    % Journal Style : pra, prb, prc, prd, pre,
               letterpaper,             % Size : a4paper, ...
               superscriptaddress,      % Affiliation (Title) : groupedaddress,
               nofootinbib,         % Footnote: footinbib, nofootinbib
               tightenlines,        % Remove additional spaces in a line
               floats,floatfix      % Floating pictures and tables
               ]{revtex4-1}

\usepackage{graphicx}% Include figure files
\usepackage{dcolumn}% Align table columns on decimal point
\usepackage{bm}% bold math
\usepackage{amsmath}
\usepackage{color}
\usepackage{enumerate}

\begin{document}

\title{Stellar models in Brane Worlds}

\author{Francisco X. Linares}
\email{linares.francisco@gmail.com}
\affiliation{Departamento de F\'isica, Universidad Sim\'on Bol\'ivar Apartado 89000, Caracas 1080A, Venezuela.}

\author{Miguel A. Garc\'{\i}a-Aspeitia} 
\email{aspeitia@fisica.uaz.edu.mx}
\affiliation{Consejo Nacional de Ciencia y Tecnolog\'ia, Av, Insurgentes Sur 1582. Colonia Cr\'edito Constructor, Del. Benito Ju\'arez C.P. 03940, M\'exico D.F. M\'exico.}
\affiliation{Unidad Acad\'emica de F\'isica, Universidad Aut\'onoma de Zacatecas, Calzada Solidaridad esquina con Paseo a la Bufa S/N C.P. 98060, Zacatecas, M\'exico.}

\author{L. Arturo Ure\~na-L\'opez}
\email{lurena@fisica.ugto.mx}
\affiliation{Departamento de F\'isica, DCI, Campus Le\'on, Universidad de 
Guanajuato, C.P. 37150, Le\'on, Guanajuato, M\'exico.}
\affiliation{Institute for Astronomy, University of Edinburgh, Royal Observatory, Edinburgh EH9 3HJ, United Kingdom.}

\begin{abstract}
We consider here a full study of stellar dynamics from the brane-world
point of view in the case of constant density and of a polytropic
fluid. We start our study cataloguing the minimal requirements to
obtain a compact object with a Schwarszchild exterior, highlighting
the low and high energy limit, the boundary conditions, and the
appropriate behavior of Weyl contributions inside and outside of the
star. Under the previous requirements we show an extensive study of
stellar behavior, starting with stars of constant density and its
extended cases with the presence of nonlocal contributions. Finally,
we focus our attention to more realistic stars with a polytropic
equation of state, specially in the case of white dwarfs, and study
their static configurations numerically. One of the main results is
that the inclusion of the Weyl functions from braneworld models allow
the existence of more compact configurations than within General
Relativity.
\end{abstract}

\draft
\date{\today}
\maketitle

\section{Introduction}
Stellar astrophysics is one of the most characteristic topics studied
by General Relativity (GR), which has helped to describe the dynamic
and evolution of stars with unprecedented
success\cite{Chandrasekhar:1985kt,*tolman1987relativity,*oppenheimer1939massive}. In
addition, the matter inside a star may be in some cases in extreme conditions
generating complicated high energy phenomena, principally in white
dwarfs, neutron stars, and others, and then a complete description of
the stellar properties requires the introduction of a particular equation of state (EoS) like in the case of
polytropes\cite{chandrasekhar1958introduction,*weinberg1972gravitation},
or even Bose-Einstein Condensates
(BEC)\cite{Mukherjee:2014kqa,*ValdezAlvarado:2012xc}.

Another interesting possibility in recent times is to consider
alternative theories of gravity and to look for their particular
signatures in stellar models, specially for some of the extreme
situations mentioned above. For instance, the authors
in\cite{Chagoya:2014fza} considered the corrections induced by a
Galileon Lagrangian in stars of constant density. Another example is given by
the so-called models of Brane-Worlds
(see\cite{PerezLorenzana:2005iv,mk} for a good review) whose main
characteristic is the existence of branes (four dimensional manifolds)
embedded in a five dimensional bulk\cite{Randall-I,*Randall-II}. This
particular geometry allows a natural extension of Einstein's
equations\cite{sms}, and introduces new degrees of freedom through quadratic terms of the energy momentum tensor, the non-local Weyl
terms, and other fields that could live in the bulk. This framework
has been used for stars with a constant density in\cite{gm}, and also
for polytropic matter with a given relationship between the quantities
arising from the non-local Weyl terms in\cite{Castro:2014xza}. It has
also been shown that the exterior solutions of these brane-stars is
not the Schwarzschild one\cite{gm,Ovalle:2013vna,*Ovalle:2013xla,*Ovalle:2014uwa}, and then
the Weyl fluids in the exterior of the stars can have a non-negligible
influence in the internal pressure and compactness of stellar
objects. More recently, the conditions for stellar stability in
brane-stars were revisited in\cite{Garcia-Aspeitia:2014pna} for a set
of hypotheses called the \emph{minimal setup}, which are consistent
with a Schwarzschild exterior. Also see\cite{Garcia-Aspeitia:2014jda}
for a study on the gravitational collapse of brane stars.

With the previous background, this paper is dedicated to the study of
the stellar equations of motion that arise from the formalism of
Brane-World theory, and the role of the Weyl functions in the regular
behavior of a stellar distribution. It is important to remark that our
main objective is to consider models of stars as realistic as
possible, and for this reason we will follow conventional wisdom in
this regard: a Schwarzschild exterior, and regularity of all functions
involved. Based on these premises, we perform numerical studies of the
so-called extended GM solution with constant density, and of a polytropic fluid.

The organization of the paper is as follows. In Sec.~\ref{Form}, we
describe the equations of stellar dynamics with branes, emphasizing
the high and low energy limits, boundary conditions, and the role
played by the Weyl functions in providing consistent and regular
solutions. Subsequently in Sec.~\ref{sec:case-const-dens}, we study the case of constant density and the extended GM solution. Also, in Sec. \ref{PBS} we study polytropic brane-stars. Finally, in Sec.~\ref{Disc} we give some conclusions and remarks.

\section{Stellar Dynamics with Branes}\label{Form}
%%%%%%%%%%%%%%%%%%%%%%%%%%%%%%%%%%%%%%%%%%%%%%%
Let us start by writing the equations of motion for an embedded brane
in a five dimensional bulk using the Randall-Sundrum II model\cite{Randall-II}. We first assume
that the Einstein equations are the gravitational equations of motion
of the 5-dimensional Universe,
\begin{equation}
  G_{AB} + \Lambda_{(5)} g_{AB} = \kappa^{2}_{(5)} T_{AB} \, .
\end{equation}
Following an appropriate computation, the modified 4dim Einstein's equation can be written as\cite{mk,sms}
\begin{equation}
  G_{\mu\nu} + \xi_{\mu\nu} + \Lambda_{(4)}g_{\mu\nu} =
  \kappa^{2}_{(4)} T_{\mu\nu} + \kappa^{4}_{(5)} \Pi_{\mu\nu} +\kappa^{2}_{(5)} F_{\mu\nu}\, , \label{Eins}
\end{equation}
where $\kappa_{(4)}$ and $\kappa_{(5)}$ are respectively the four and
five- dimensional coupling constants, which are related one to each
other in the form: $\kappa^{2}_{(4)}=8\pi G_{N}=\kappa^{4}_{(5)} \lambda/6$, $\lambda$ is defined as the brane tension, and $G_{N}$ is Newton's constant. For purposes of simplicity, we will not consider bulk matter, which translates into $F_{\mu\nu}=0$, and discard the presence of the four-dimensional cosmological constant, $\Lambda_{(4)}=0$, as we do not expect it to have any important effect at astrophysical scales (for a recent discussion about it see\cite{Pavlidou:2013zha}). Additionally, we will neglect any nonlocal energy flux, which is allowed by the static spherically symmetric solutions we will study below\cite{mk}.

In the case of a perfect fluid, the energy-momentum $T_{\mu\nu}$ tensor, the
quadratic energy-momentum tensor $\Pi_{\mu\nu}$, and the Weyl
$\xi_{\mu\nu}$ can be written as:
\begin{subequations}
  \begin{eqnarray}
    \label{Tmunu}
    T_{\mu\nu} &=& \rho u_{\mu}u_{\nu}+ph_{\mu\nu} \, , \\
    \label{Pimunu}
    \Pi_{\mu\nu} &=& \frac{1}{12}\rho[\rho
    u_{\mu}u_{\nu}+(\rho+2p)h_{\mu\nu}] \, , \\ \label{ximunu}
    \xi_{\mu\nu} &=& - \frac{6}{\kappa_{(4)}^{2}\lambda} \left[
      \mathcal{U}u_{\mu}u_{\nu} +\mathcal{P} r_{\mu}r_{\nu} +
      \frac{\mathcal{U} - \mathcal{P}}{3}h_{\mu\nu} \right]\, ,
  \end{eqnarray}
\end{subequations}
where $p=p(r)$ and $\rho=\rho(r)$ are respectively, the pressure and
density of the stellar matter of interest, $\mathcal{U}$ is the
nonlocal energy density, $\mathcal{P}$ is the nonlocal anisotropic
stress scalar, $u_{\alpha}$ is the fluid four-velocity, that also
satisfies the condition $g_{\mu\nu}u^{\mu}u^{\nu}=-1$, and
$h_{\mu\nu} = g_{\mu\nu} + u_{\mu}u_{\nu}$ is orthogonal to
$u_{\mu}$. Under the assumption of spherical symmetry, the metric can be written as:
\begin{equation}
{ds}^{2}=-B(r){dt}^{2}+A(r){dr}^{2}+r^{2}({d\theta}^{2}+{\sin}^{2}\theta{d\varphi}^{2}) \, .
\label{metric}
\end{equation}
The equations of motion for stellar structure then are\cite{gm,Garcia-Aspeitia:2014pna}:
\begin{subequations}
  \label{eq:1}
  \begin{eqnarray}
    \mathcal{M}^\prime &=& 4\pi r^2 \rho_{eff} \, , \label{eq:1a} \\
    p^{\prime} &=& -\frac{1}{2} \frac{B^{\prime}}{B} (p+\rho) \, , \label{eq:1b} \\ 
    \mathcal{V}^{\prime} + 3 \mathcal{N}^{\prime} &=& - \frac{B^{\prime}}{B} (2\mathcal{V} + 3\mathcal{N}) - \frac{9}{r}\mathcal{N} - 3(\rho+p)\rho^{\prime} \, , \label{eq:1d} \\
    \frac{B^{\prime}}{B} &=& \frac{2G_N}{r^2} \left( \frac{ 4\pi r^3 p_{eff} + \mathcal{M} }{1 - 2G_N\mathcal{M}/r} \right) \, , \label{eq:1c}
  \end{eqnarray}
\end{subequations}
where a prime indicates derivative with respect to $r$. We have also defined $\mathcal{V} = 6 \mathcal{U} /\kappa^{4}_{(4)}$, $\mathcal{N} = 4\mathcal{P} /\kappa^{4}_{(4)}$, and $A(r) = [1-2G_{_{N}} \mathcal{M}(r)/r]^{-1}$, whereas $p_{eff}$ and $\rho_{eff}$ are explicitly given by:
\begin{subequations}
  \label{eq:2}
  \begin{eqnarray}
    p_{eff} &=& p \left( 1 +\frac{\rho}{\lambda} \right) + \frac{\rho^{2}}{2\lambda} + \frac{\mathcal{V}}{3\lambda} + \frac{\mathcal{N}}{\lambda} \, , \label{peff} \\
    \rho_{eff} &=& \rho \left( 1 +\frac{\rho}{2\lambda} \right) +\frac{\mathcal{V}}{\lambda} \, . \label{reff}
  \end{eqnarray}
\end{subequations}
Now, we are in position of analyze the following important points.

\subsection{Numerical analysis} \label{CD}
%%%%%%%%%%%%%%%%%%%%%%%%%%%%%%%%%%%%%%%%%%%%%%
In order to have a numerical solution of the equations of motion, we choose the following dimensionless variables:
\begin{subequations}
  \label{eq:10}
\begin{eqnarray}
  x &=& \sqrt{G_NM/R} \, (r/R) \, , \, \bar{\rho} = \rho/\langle \rho_{eff} \rangle \, , \, \bar{p} = p/\langle \rho_{eff} \rangle \, ,  \label{eq:10a} \\
  \bar{\lambda} &=& \lambda/\langle \rho_{eff} \rangle \, , \, \bar{\mathcal{V}} = \mathcal{V}/\langle \rho_{eff} \rangle^{2} \, , \, \bar{\mathcal{N}} = \mathcal{N}/\langle \rho_{eff} \rangle^{2} \, , \label{eq:10b}
\end{eqnarray}
\end{subequations}
for which Eqs.~\eqref{eq:1} read
\begin{subequations}
  \label{eq:16}
  \begin{eqnarray}
    \bar{\mathcal{M}}^\prime &=&  x^2 \bar{\rho}_{eff} \, \label{eq:16a} \, , \\
   \bar{p}^\prime &=&  - \frac{3}{x^2} \left( \frac{x^3 \bar{p}_{eff} + \bar{\mathcal{M}}}{1 - 6 \bar{\mathcal{M}}/x} \right) ( \bar{p} + \bar{\rho} ) \, ,   \label{eq:16b} \\
    \bar{\mathcal{V}}^{\prime} + 3 \bar{\mathcal{N}}^{\prime} &=& - \frac{6}{x^2} \left( \frac{x^3 \bar{p}_{eff} + \bar{\mathcal{M}}}{1 - 6 \bar{\mathcal{M}}/x} \right) (2\bar{\mathcal{V}} + 3\bar{\mathcal{N}}) \nonumber \\
&& - \frac{9}{x} \bar{\mathcal{N}} - 3(\bar{\rho} + \bar{p}) \bar{\rho}^{\prime} \, , \label{eq:16c}
  \end{eqnarray}
\end{subequations}
where now a prime denotes derivatives with respect to $x$, the mean effective density is just given by $\langle \rho_{eff} \rangle = 3M/4\pi R^3$, and also
\begin{subequations}
\label{eq:17}
\begin{eqnarray}
\bar{\rho}_{eff}  &=& \bar{\rho} \left( 1 + \frac{\bar{\rho}}{2\bar{\lambda}} \right) + \frac{\bar{\mathcal{V}}}{\bar{\lambda}}  \, , \label{eq:17a} \\
\bar{p}_{eff} &=& \bar{p} \left(1 + \frac{\bar{\rho}}{\bar{\lambda}} \right) + \frac{\bar{\rho}^{2}}{2\bar{\lambda}} + \frac{\bar{\mathcal{V}}}{3\bar{\lambda}}+\frac{\bar{\mathcal{N}}}{\bar{\lambda}} \, . \label{eq:17b}
\end{eqnarray}
\end{subequations}
Note that the ratio $\rho/\lambda$ is invariant under the change of variables, $\bar{\rho}/\bar{\lambda}=\rho/\lambda$, and then we will omit the bar whenever the ratio appears in the equations of motion.

\subsection{High and low energy limits}
\label{sec:high-low-energy}
There are two very clear limiting expressions of the equations of motion in terms of normalized brane ratio $\bar{\lambda}$, as the latter represents the energy ratio of the brane tension with respect to the mean energy density of the compact star of interest, see Eq.~\eqref{eq:10b}. It is usually assumed that the brane corrections are measured in terms of the absolute value of the brane tension $\lambda$, but in our study we find the brane ratio $\bar{\lambda} = \lambda/\langle \rho_{eff} \rangle$ to have a more meaningful character for compact objects in general.

Under this line of reasoning, we first present the low energy limit of
the equations of motion, represented by the operation $\bar{\lambda}
\to \infty$, under which Eqs.~\eqref{eq:16} become the usual
Tolman-Oppenheimer-Volkoff (TOV) equations of GR\cite{weinberg1972gravitation,tolman1987relativity,oppenheimer1939massive}:
\begin{subequations}
  \label{eq:30}
  \begin{eqnarray}
    \bar{\mathcal{M}}^\prime &=&  x^2 \bar{\rho} \, \label{eq:30a} \, , \\
    \bar{p}^\prime &=&  - \frac{3}{x^2} \left( \frac{x^3 \bar{p} + \bar{\mathcal{M}}}{1 - 6 \bar{\mathcal{M}}/x} \right) ( \bar{p} + \bar{\rho} ) \, , \label{eq:30b}
  \end{eqnarray}
\end{subequations}
where the effective pressure and density are directly represented by their normalized physical values. 

We have called it the low energy limit because we are assuming that the mean density of the star is much lower than the brane tension $\lambda$, so that any brane corrections in the equations of motion are highly suppressed by the brane energy scale. We cannot say here whether the brane tension is at a very energy scale, or it is just that the star density is not high enough. In strict sense, Eq.~\eqref{eq:16c} can still be considered for the integration of the Weyl functions, but their values will not make any difference in the final integration of the physical variables.

There is also the high energy limit of the equations of motion represented by $\bar{\lambda} \to 0$, for which the effective density and pressure read
\begin{subequations}
\label{eq:31}
\begin{eqnarray}
\bar{\rho}_{eff}  & \simeq & \frac{\bar{\rho}^2}{2\bar{\lambda}} + \frac{\bar{\mathcal{V}}}{\bar{\lambda}}  \, , \label{eq:31a} \\
\bar{p}_{eff} & \simeq & \frac{\rho}{2\lambda}(2\bar{p}+\bar{\rho}) + \frac{\bar{\mathcal{V}}}{3\bar{\lambda}} + \frac{\bar{\mathcal{N}}}{\bar{\lambda}} \, . \label{eq:31b}
\end{eqnarray}
\end{subequations}

We can see that there is an overall factor of $\bar{\lambda}$ in the
above expressions~\eqref{eq:31}, which will also appear as such in
Eqs.~\eqref{eq:16} in the high energy limit. The brane ratio can then
be absorbed in the equations of motion by means of the following
change of variables: $\bar{\mathcal{M}} \to \bar{\mathcal{M}}
\bar{\lambda}^{1/2}$, and $x \to x \bar{\lambda}^{1/2}$, and then we
finally find the equations of motion for the high energy limit:
\begin{subequations}
\label{eq:32}
  \begin{eqnarray}
    \bar{\mathcal{M}}^\prime &=&  \frac{x^2}{2} \left( \bar{\rho}^2 + 2 \bar{\mathcal{V}} \right) \, \label{eq:32a} \, , \\
   \bar{p}^\prime &=&  - \frac{3}{x^2} \left[ \frac{x^3 (\bar{p} \bar{\rho} + \bar{\rho}^{2}/2 + \bar{\mathcal{V}}/3 + \bar{\mathcal{N}}) + \bar{\mathcal{M}}}{1 - 6 \bar{\mathcal{M}}/x} \right] \times\nonumber\\&&( \bar{p} + \bar{\rho} ) \, ,   \label{eq:16b} \\
    \bar{\mathcal{V}}^{\prime} + 3 \bar{\mathcal{N}}^{\prime} &=& - \frac{6}{x^2} \left( \frac{x^3 (\bar{p} \bar{\rho} + \bar{\rho}^{2}/2 + \bar{\mathcal{V}}/3 + \bar{\mathcal{N}}) + \bar{\mathcal{M}}}{1 - 6 \bar{\mathcal{M}}/x} \right) \times\nonumber\\&&(2\bar{\mathcal{V}} + 3\bar{\mathcal{N}}) - \frac{9}{x} \bar{\mathcal{N}} - 3(\bar{\rho} + \bar{p}) \bar{\rho}^{\prime} \, .
  \end{eqnarray}
\end{subequations}

In contrast to the TOV equations of GR in~\eqref{eq:30}, we shall
call this case the high energy limit because the mean density is much
larger than the brane tension, even if we cannot say whether this is
because the brane tension attains a very small energy value, or it is
just that the star has such a large density that the latter surpasses the energy scale of the brane tension.

\subsection{Boundary conditions \label{sec:boundary-conditions-}}
The change of variables~\eqref{eq:10} is very appropriate to explore the solutions of the TOV equations~\eqref{eq:16}, as all physical quantities involved are normalized in terms of two important observables in stellar astrophysics, which are the mass $M$ and the radius $R$ of the star. Furthermore, these two parameters appear in the single combination $G_N M/R$ that represents the compactness of the star. For instance, the interior range of the new radial variable is $x = [0, \sqrt{G_N M/R}]$, which means that the surface of the star is located at $x(R) \equiv X = \sqrt{G_N M/R}$. Also, the new mass function changes to: 
\begin{equation}
  \label{eq:18}
  \bar{\mathcal{M}}(x)  = \frac{1}{3} \left( \frac{G_N M}{R} \right)^{3/2} \frac{\mathcal{M}(r)}{M} \, ,
\end{equation}
and then the total mass is $\bar{\mathcal{M}}(X) = (1/3) (G_N M /R)^{3/2}$. In other words, the compactness of the star will determine the mass and size of the numerical solutions.

The equations of motion will be integrated from the center up to the surface of the star defined by the condition $p(X)=0$; the latter only refers to the physical pressure, and we will take it as a reasonable physical assumption even though it is not necessarily required in the case of brane stars. Finally, at the center of the star we will also assume that $\bar{\mathcal{M}} \to 0$ as $x\to 0$, so that there is not a discontinuity of the different quantities in the center of the star, and the central value of the pressure (or any other related quantity) will be set as a free parameter that will characterize the numerical solutions. 

Even though we will not consider exterior solutions, we must anyway take into account the information provided by the Israel-Darmois (ID) matching condition, which for the case under study can be written as\cite{gm}:
\begin{equation}
  \label{eq:9}
  (3/2) \bar{\rho}^2(X) + \bar{\mathcal{V}}^-(X) + 3 \bar{\mathcal{N}}^-(X) = \bar{\mathcal{V}}^+(X) + 3 \bar{\mathcal{N}}^+(X) \, ,
\end{equation}
where the superscript $-(+)$ denotes the interior (exterior) values of the different quantities at the surface of the star, and we also assumed that $\bar{\rho}(x> X) =0$.

A desirable property we want in our solutions is a \emph{Schwarzschild
  exterior}, which can be easily accomplished under the boundary
conditions $\bar{\mathcal{V}}^+(X) = 0 =\bar{\mathcal{N}}^+(X)$, as
for them the simplest solution that arises from Eq.~\eqref{eq:16c} is
the trivial one: $\bar{\mathcal{V}}(x \geq X) = 0 =\bar{\mathcal{N}}(x
\geq X)$. Thus, for the purposes of this paper, we will refer
hereafter to the restricted ID matching condition given by:
\begin{equation}
  \label{eq:28}
    (3/2) \bar{\rho}^2(X) + \bar{\mathcal{V}}^-(X) + 3 \bar{\mathcal{N}}^-(X) = 0 \, .
\end{equation}

For completeness, we just note that the exterior solutions of the metric functions are given by the well known expressions $B(r) = A^{-1}(r) = 1 - 2G_N M/r$. In addition, it can be shown from Eq.~\eqref{eq:1c} that the interior solution of the lapse function, in terms of the normalized variables~\eqref{eq:10}, is given by:
\begin{equation}
  \label{eq:29}
  \frac{B(x)}{1 - 2 G_N M/R} = \exp \left[ - \int_x^X dx \frac{6}{x^2} \left( \frac{x^3 \bar{p}_{eff} + \bar{\mathcal{M}}}{1 - 6 \bar{\mathcal{M}}/x} \right) \right] \, ,
\end{equation}
and then we will not solve it explicitly in any of the cases presented below.

The numerical recipe described above will be applied to different
cases and configurations in Secs.~\ref{sec:case-const-dens}
and~\ref{PBS}. The results that will be obtained will have a universal
character, as they will not depend upon the particular values of the
mass and radius of a given star, but they will represent general
classes of stars according to their common compactness $G_N M/R$. This
will allow us to reach wide general conclusions about the physical
properties of the different configurations by means of numerical
methods.

\subsection{Weyl functions}
\label{sec:weyl-functions}
It must be noticed that the interior solutions cannot evade the presence of the Weyl terms even if the exterior solution is Schwarzschild. For example, let us put by hand that $\bar{\mathcal{N}}(x) \equiv 0$. If the density is constant $\bar{\rho}(X) \neq 0$, the ID matching condition~\eqref{eq:28} implies that $\bar{\mathcal{V}}^-(X) = -(3/2) \bar{\rho}^2(X)$, and then the full solution must be\cite{gm}: 
\begin{equation}
  \label{eq:8}
  \mathcal{V}(x < X) = -(3/2) \rho^2 (1+p/\rho)^4 \, .
\end{equation}
The full consequences of this nonlocal energy density $\mathcal{V}$ are explored in Sec.~\ref{sec:extended-gm-solution-1} below.

Even the condition of a Schwarzschild exterior together with $\bar{\rho}(X)=0$ do not directly imply that the Weyl functions must vanish in the interior, as it can be shown\cite{Garcia-Aspeitia:2014pna} that in such a case Eq.~\eqref{eq:16c} must have the following solution:
\begin{equation}
\label{eq:19}
    \mathcal{V} (x < X) = \frac{3}{B^2(x)} \int^X_x B^2 (\bar{\rho} + \bar{p}) \bar{\rho}^\prime \, dx \, ,
\end{equation}
which accomplishes the boundary condition $\mathcal{V}^{-}(X)=0$ and is also regular at $x=0$. This is particularly important for all cases in which the density is not constant, as we shall see below for the polytropes in Sec.~\ref{PBS}. 

In the opposite case when $\bar{\mathcal{V}}(x) \equiv 0$ and the
density is constant, the ID matching condition~\eqref{eq:28} implies
that at the surface of the star: $\bar{\mathcal{N}}^-(X) = -(1/2) \bar{\rho}^2(X)$, and then Eq.~\eqref{eq:16c} integrates into
\begin{equation}
  \label{eq:25}
  \mathcal{N}(x < X) = -\frac{1}{2} \left( \frac{X}{x} \right)^3 (p + \rho)^2 \, .
\end{equation}
Needless to say, this solution diverges at the center of the star and cannot be considered as a useful interior solution. That is, \emph{in the case of constant density there is not a regular interior solution with the only presence of the nonlocal anisotropic stress $\mathcal{N}$}.

There is though a non-divergent interior solution of $\mathcal{N}$ if we drop the condition of constant density, which is:
\begin{equation}
\label{eq:19-0}
\mathcal{N}(x <X) = \frac{1}{B(x)x^3} \int_{0}^{x} B x^{3} (\bar{\rho} + \bar{p}) \bar{\rho}^{\prime} dx \, .
\end{equation}
But the ID matching condition~\eqref{eq:28} now indicates that at the
surface of the star we must have $\bar{\rho}(X) \neq 0$, or either
give up the Schwarzschild exterior. In consequence, \emph{the only
  interior solution of the nonlocal anisotropic stress under the
  conditions of a Schwarzschild exterior, and non-constant density
  with $\bar{\rho}(X) = 0$, which are the conditions we expect to have
  in realistic stars, is the trivial one: $\mathcal{N}(x) \equiv 0$}
(see also \cite{Garcia-Aspeitia:2014pna}).

There are other possibilities that have been explored in the specialized literature, like for instance a relationship between the Weyl functions in the form $\mathcal{N}= \sigma \mathcal{V}$, where $\sigma$ is a constant parameter\cite{Castro:2014xza}. Clearly, the solutions~\eqref{eq:8} and~\eqref{eq:19} are special cases for which $\sigma =0$. In the general case, Eq.~\eqref{eq:16c} can be written as:
\begin{equation}
  \label{eq:6}
  (1+3\sigma) \bar{\mathcal{V}}^{\prime} = - \frac{B^{\prime}}{B} (2 + 3 \sigma) \bar{\mathcal{V}} - \frac{9}{x} \sigma \bar{\mathcal{V}} - 3(\bar{\rho} + \bar{p}) \bar{\rho}^{\prime} \, ,
\end{equation}
as long as $\sigma \neq -1/3$. If the density is constant, then there is a solution which is similar to Eq.~\eqref{eq:25}:
\begin{equation}
  \label{eq:26}
  \bar{\mathcal{V}} = C \left[ (p + \rho)^{2(2+3\sigma)} x^{-9\sigma} \right]^{1/(1+3\sigma)} \, ,
\end{equation}
where $C$ is an integration constant that could be determined with the help of the ID matching condition~\eqref{eq:28}. However, this solution is not appropriate for the interior of the star because, as it happened too for Eq.~\eqref{eq:25}, it diverges in the center of the star ($x=0$).

We can also consider the case of non-constant density, in which case we find a similar solution to Eq.~\eqref{eq:19-0}:
\begin{eqnarray}
  \mathcal{N}(x <X) &=& \left[ B(x)^{-2(2+3\sigma)} x^{-9\sigma} \right]^{1/(1+3\sigma)} \times \nonumber \\
&& \int_{0}^{x} \left[ B^{2(2+3\sigma)} x^{9\sigma} \right]^{1/(1+3\sigma)} (\bar{\rho} + \bar{p}) \bar{\rho}^{\prime} dx \, . \label{eq:27}
\end{eqnarray}
Even though this solution is well behaved in the interior of the star, it needs a non-trivial boundary condition at the surface as dictated by the ID matching condition~\eqref{eq:28}, and for that we require either to have $\bar{\rho}(X) \neq 0$, or to give up the Schwarzschild exterior.

We see that the imposition of a Schwarzschild exterior has strong consequences for the interior solutions of the Weyl functions, mostly because it is difficult in general to find for them a well behaved interior solution. The most problematic case is that of the anisotropic stress function $\mathcal{N}$, and for this reason we will not take it into account as part of the brane gravitational corrections, but assume that the latter are only given by the quadratic corrections of the density $\bar{\rho}^2$ and the nonlocal energy density $\mathcal{V}$.

\section{The case of constant density \label{sec:case-const-dens}}
One of the simplest possibilities of star modes is that of constant
density $\rho$, which can be solved under different gravitational
schemes. In this section we will work out such a case within the
braneworld scheme and explain the additional physical and boundary conditions that may be needed in order to reach well posed numerical solutions.

\subsection{The case of the Germani-Maartens solution of brane stars \label{sec:case-germ-maart}}
To start with we consider here the GM interior solution, which was
thoroughly studied in\cite{gm}, and that does not take into account
corrections induced by the Weyl terms: $\bar{\mathcal{V}} = 0 =
\bar{\mathcal{N}}$. The modified TOV equations are given again by
Eqs.~\eqref{eq:1a} and~\eqref{eq:1b} with the following
identifications: $\rho_{eff} = \rho (1+\rho/2\lambda)$, and $p_{eff} =
p (1+ \rho/\lambda)+\rho^2/2\lambda$.

Because the density is constant, we find, in terms of the
variables in~\eqref{eq:10}, that $\langle \rho_{eff} \rangle= \rho (1+ \rho/2\lambda)$ and then $\bar{\rho}_{eff} = 1$. Likewise, we find that $\bar{\rho} = (1+\rho/2\lambda)^{-1}$, and then its value is directly determined by the ratio $\rho/\lambda$. Notice that $\bar{\rho} \leq 1$, and that we recover $\bar{\rho} =1$ in the GR limit $\rho/\lambda \to 0$. The boundary conditions depend upon the compactness of the star $G_N M/R$, as in the case of GR, but also upon the ratio $\rho/\lambda$, as expected in brane models. The exact solution of the pressure function is\cite{gm}:
\begin{equation}
\frac{\bar{p}}{\bar{\rho}}=\frac{\sqrt{1-6\bar{\mathcal{M}}/X} - \sqrt{1-6\bar{\mathcal{M}}x^{2}/X^{3}} }{  \sqrt{1-6\bar{\mathcal{M}}x^{2}/X^{3}} - 3\zeta^{-1} \sqrt{1-6\bar{\mathcal{M}}/X} },
\end{equation}
where $\zeta \equiv (1+2\rho/\lambda)/(1+ \rho/\lambda)$. The brane
ratio $\rho/\lambda$ lowers the maximum value of the compactness of
the star, and the numerical solutions satisfies the analytic bound
found from the exact GM solution: $G_N M/R = (1/2) (1 - \zeta^2/9)$. The GR limit is obtained when $\rho/\lambda \to 0$: $G_N M/R \leq
4/9$, whereas in the opposite direction $\rho/\lambda \to \infty$ we
obtain: $G_N M/R \leq 5/18$. 

According to the discussion in Sec.~\ref{sec:weyl-functions}, the GM
solution cannot be matched to a Schwarzschild exterior, and for that
reason it is usually assumed that other exterior solutions with the presence of the Weyl function must be the correct ones for brane stars. However, it has been recently shown\cite{Garcia-Aspeitia:2014pna}, under very general conditions, that the GM solution plays also the role of being the limiting case of realistic stars when brane corrections are considered, and then gives an upper bound in the compactness of stars with both brane corrections and a Schwarzschild exterior.

\subsection{The extended GM solution \label{sec:extended-gm-solution-1}}
We will now review the interior brane solution with constant density, a
Schwarzschild exterior, and a non-null Weyl term $\mathcal{V}$. This
cases was also briefly considered in\cite{gm}, but lacks an analytical solution. The equations of motion are again~\eqref{eq:16} with the following expressions for the effective density and pressure:
\begin{subequations}
\label{eq:7}
\begin{eqnarray}
\bar{\rho}_{eff}  &=& \bar{\rho} \left( 1 + \frac{\rho}{2\lambda} \right) - \frac{3}{2} \bar{\rho}\frac{\rho}{\lambda}  (1 + \bar{p}/\bar{\rho})^4 \, , \label{eq:7a}\\
\bar{p}_{eff} &=& \bar{p} \left(1 + \frac{\rho}{\lambda} \right) + \frac{\bar{\rho}}{2} \frac{\rho}{\lambda} - \frac{\bar{\rho}}{2} \frac{\rho}{\lambda}  (1 + \bar{p}/\bar{\rho})^4 \, . \label{eq:7b}
\end{eqnarray}
\end{subequations}
Here we have taken into account that the nonlocal energy density is
given by Eq.~\eqref{eq:8}. As it can be seen in Eqs.~\eqref{eq:7}, there are negative contributions in both the effective density and pressure originated from the presence of the Weyl nonlocal energy, and the solutions now depend upon three separate parameters: the constant density $\rho$, the brane ratio $\rho/\lambda$, and the compactness of the star $G_N M/R$. 

As we expect to have $p(x) > 0$, then the effective density must be an
increasing function, $\rho_{eff} (x) \leq \rho_{eff} (X)$, which may
even attain negative values at the interior points where the pressure
is largest. Moreover, we also infer from this information that
$\langle \rho_{eff} \rangle < \rho_{eff} (X) = \rho (1-
\rho/2\lambda)$, and then we expect that in general $\bar{\rho} > (1-
\rho/2\lambda)^{-1}$. In contrast to the GM case above, $\bar{\rho}$
cannot be given a fixed value beforehand and becomes a variable that
must be adjusted appropriately so that the numerical solutions
accomplish all boundary conditions. This time, however, the GR limit
$\bar{\rho}=1$ is a lower bound as $\rho/\lambda \to 0$, which is an
early indication that the known GR upper bound on the star compactness
could in principle be surpassed by the new solutions.

The equations of motion are more easily solved if we take the following change of variables:
\begin{equation}
\label{eq:23}
  x \to x \bar{\rho}^{-1/2} \, , \; \bar{\mathcal{M}} \to \bar{\mathcal{M}} \bar{\rho}^{-1/2} \, , \; w \equiv p/\rho \, ,
\end{equation}
where $w$ is the EoS, and then the Eqs.~\eqref{eq:16} become:
\begin{subequations}
\label{eq:24}
  \begin{eqnarray}
    \bar{\mathcal{M}}^\prime &=&  x^2 \left[ 1 + \frac{\rho}{2\lambda} - \frac{3}{2} \frac{\rho}{\lambda}  (1 + w )^4 \right] \, \label{eq:24a} \, , \\
    w^\prime &=&  - \frac{3}{x^2} \left( \frac{x^3 w_{eff} + \bar{\mathcal{M}}}{1 - 6 \bar{\mathcal{M}}/x} \right) ( 1 + w ) \, ,   \label{eq:24b} \\
w_{eff} &=& \frac{\bar{p}_{eff}}{\bar{\rho}} = w \left(1 + \frac{\rho}{\lambda} \right) + \frac{\rho}{2 \lambda} \left[ 1 - (1 + w)^4 \right] \, . \nonumber
  \end{eqnarray}
\end{subequations}
The only free parameter that appears explicitly in Eqs.~\eqref{eq:24} is the brane ratio $\rho/\lambda$. Moreover, the outer boundary conditions, see for instance Eq.~\eqref{eq:18}, must be adjusted to the values $X = (G_NM/R)^{1/2} \bar{\rho}^{1/2}$ and $\bar{\mathcal{M}}(X) = (G_N M/R)^{3/2} \bar{\rho}^{1/2}/3$.

\begin{figure}[htp!]
\includegraphics[scale=0.7]{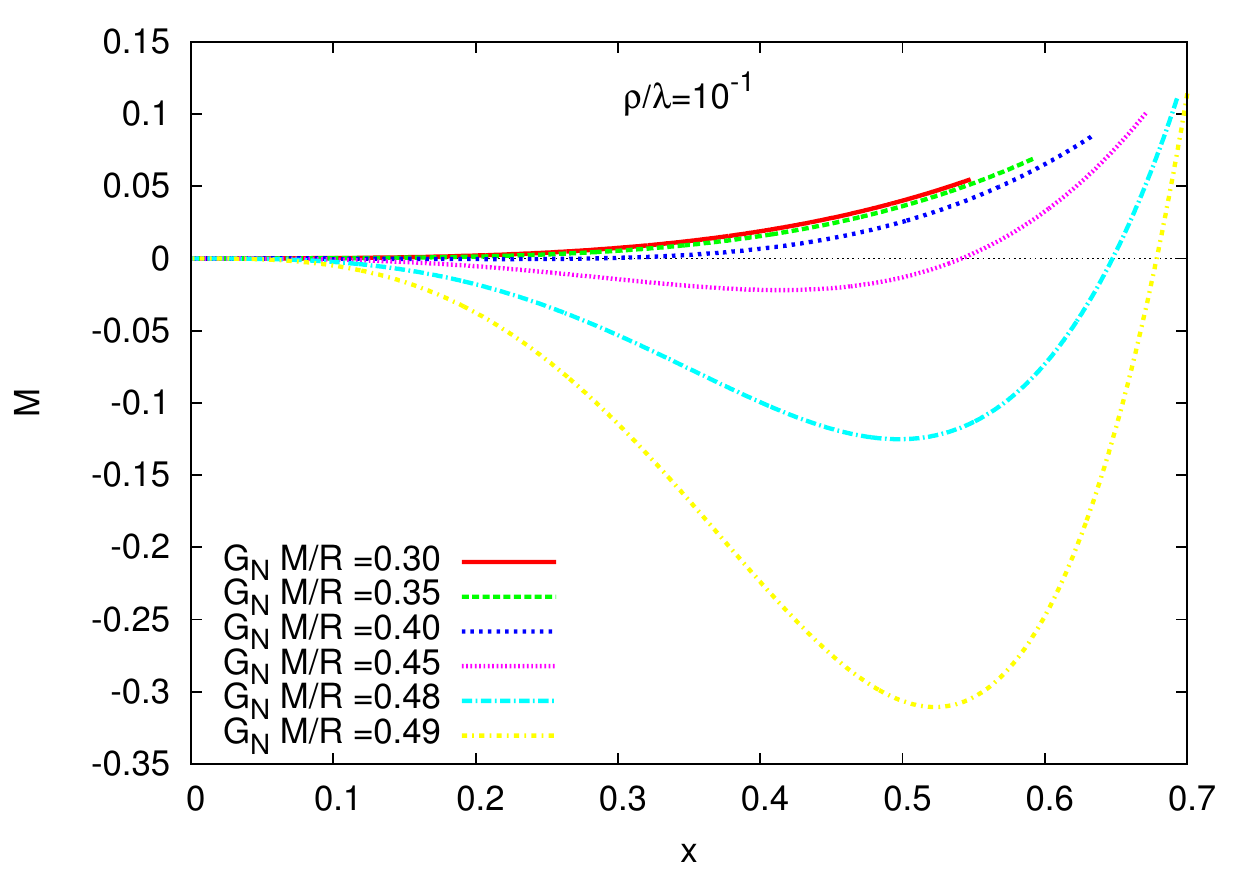}
\includegraphics[scale=0.7]{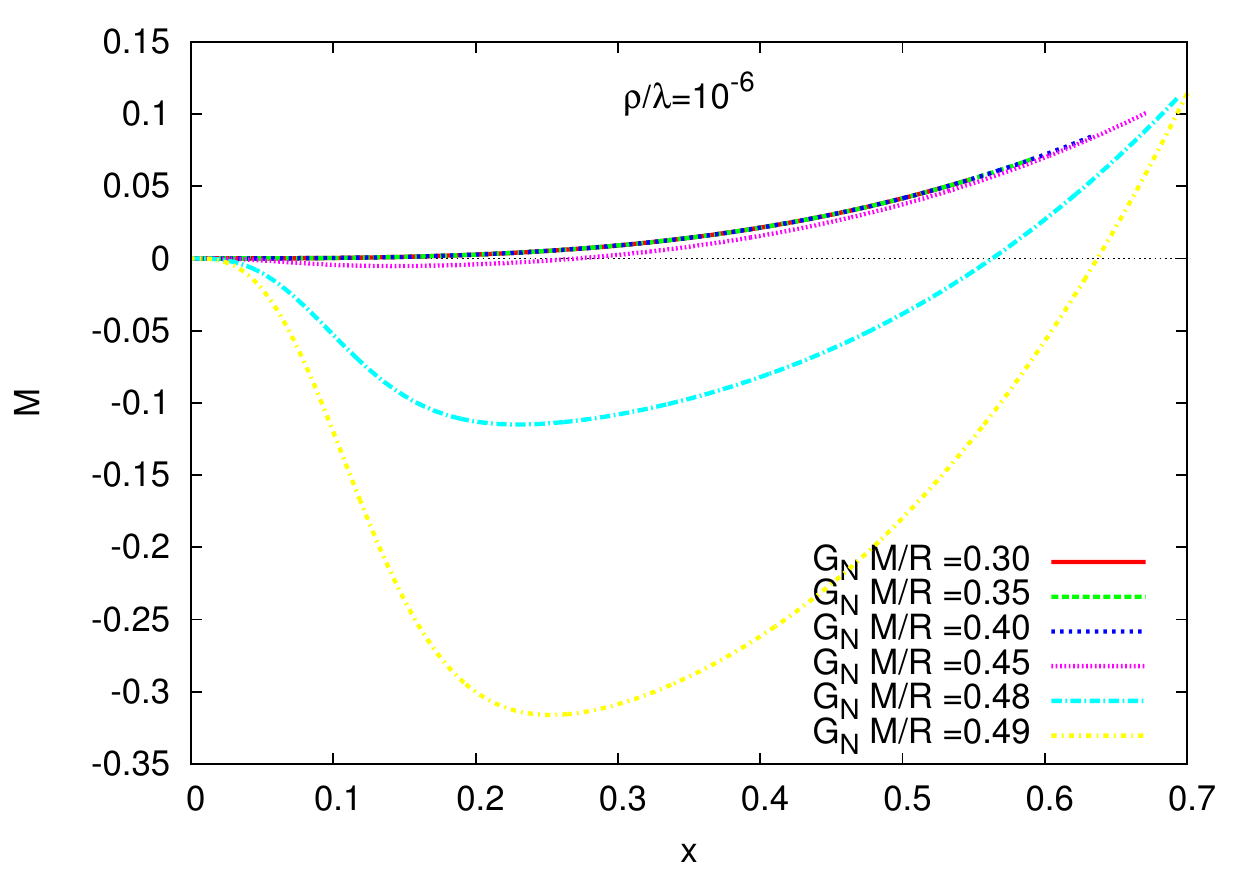}
\caption{The profile of the integrated mass $\mathcal{M}(x)$ corresponding to the extended GM solution with ratios $\rho/\lambda=10^{-1}$ (Top) and $\rho/\lambda=10^{-6}$ (Bottom). We can see that the extended GM solution allows the existence of stars with a compactness beyond the GR limit but below the extreme Schwarzshild limit $G_NM/R < 1/2$. Notice that one reason for that is that the mass function can acquire negative values in the interior of the star for the most compact cases. See the text for more details.\label{fig:GMWconstant}}
\end{figure}

Examples of the numerical solutions allowed by Eqs.~\eqref{eq:24} are
shown in Fig.~\ref{fig:GMWconstant} for the brane ratios $\rho/\lambda
=10^{-1},10^{-6}$, where it is confirmed that there are numerical
solutions well beyond the GR limit of $G_N M/R \leq 4/9$. We only
considered cases in which the star has an overall positive mass, for
which it must also have a positive density at its surface. The latter
can be translated into the condition $\bar{\rho}_{eff}(X) >0$, and
then from Eq.~\eqref{eq:7a} we find the constraint $\rho/\lambda < 1$.

There are two main reasons for the surpass of the GR limit. The first
one is that the extra free parameter $\bar{\rho}$ is only bounded from
below, and then it is at our disposal to find numerical solutions that
can surpass the GR limit for any given value of the brane ratio
$\rho/\lambda$. Correspondingly, the second reason is that the
effective density $\bar{\rho}_{eff}$ is an increasing function that
can become as negative as necessary in the interior of the
star. Actually, as far as the numerical experiments are concerned, the
only true limit that could be found for the numerical solutions is the
Schwarzschild one $G_NM/R < 1/2$.

%%%%%%%%%%%%%%%%%%%%%%%%%%%%%%%%%
\section{Polytropic brane stars} \label{PBS}
%%%%%%%%%%%%%%%%%%%%%%%%%%%%%%%%%
In this section we study brane stars with a polytropic fluid and an EoS in
the form $p(r)=K\rho^{\gamma}(r)$. Here, $K$ is the polytropic constant, and $\gamma$ is the polytropic exponent, which can be written in terms of the polytropic index $n$ as $\gamma\equiv (n+1)/n$. For example, white dwarfs can be modeled by the polytropic index $n=3$, and neutron stars by polytropes with an index in the range $n=0.5-1$\cite{chandrasekhar1935highly}.

The equations of motion~\eqref{eq:16} can be simplified if we follow
the usual recipe for polytropes and make the following change of
variable for the density: $\bar{\rho} = \theta^n$, where $n$ is the
polytropic index defined above. For the reasons explained in
Sec.~\ref{sec:weyl-functions}, we set $\bar{\mathcal{N}}=0$. Eqs.~\eqref{eq:16} are then written in the form:
\begin{subequations}
\label{eq:13}
  \begin{eqnarray}
    \bar{\mathcal{M}^\prime} &=& x^{2} \left[ \theta^n \left( 1 + \frac{\theta^n}{2\bar{\lambda}} \right) + \frac{\bar{\mathcal{V}}}{\bar{\lambda}} \right] \ , \label{eq:13a} \\
    \theta^\prime &=& - \frac{3}{x^2} \left( \frac{x^3 \bar{p}_{eff} + \bar{\mathcal{M}}}{1 - 6 \bar{\mathcal{M}}/x} \right) \frac{\left( 1+ \bar{K} \theta \right)}{\bar{K} (n+1)} \, , \label{eq:13b} \\
    \bar{\mathcal{V}}^\prime &=& - \frac{12}{x^2} \left( \frac{x^3 \bar{p}_{eff} + \bar{\mathcal{M}}}{1 - 6 \bar{\mathcal{M}}/x} \right) \bar{\mathcal{V}} \nonumber \\
&& + \frac{9}{x^2} \left( \frac{x^3 \bar{p}_{eff} + \bar{\mathcal{M}}}{1 - 6 \bar{\mathcal{M}}/x} \right) \frac{ n ( 1 + \bar{K} \theta )^2}{\bar{K}(n+1) } \theta^{2n-1} \, , \label{eq:13c}
  \end{eqnarray}
\end{subequations}
and the effective pressure~\eqref{eq:17b} now reads
\begin{equation}
\label{eq:14}
  \bar{p}_{eff} = \bar{K} \theta^{n+1} \left(1 + \frac{\theta^n}{\bar{\lambda}}  \right) + \frac{\theta^{2n}}{2\bar{\lambda}} + \frac{\bar{\mathcal{V}}}{3\bar{\lambda}} \, .
\end{equation}
It must be stressed out that in our case the density parameter
$\theta$ gives an indication of the values of the density $\rho$ with
respect to the mean value of the effective density $\langle \rho_{eff}
\rangle$, given by the dimensionless density $\bar{\rho}$, in contrast
to the standard case in which the value of reference is the density at
the center of the star $\rho(0)$.

It can also be shown that, again like in the standard case of
polytropes, the polytropic coefficient $\bar{K}$ is a redundant
constant and can be hidden in the equations of motion. If we further
consider the following change of variables:
\begin{subequations}
  \label{eq:20}
\begin{eqnarray}
  x \to x \bar{K}^{n/2} \, , \; \theta \to \theta \bar{K}^{-1} \, , \; \bar{\mathcal{M}} \to \bar{\mathcal{M}} \bar{K}^{n/2} \, , \\
\bar{p}_{eff}\to\bar{p}_{eff}K^{-n} \, , \; \bar{\lambda} \to \bar{\lambda} \bar{K}^{-n} \, , \; \bar{\mathcal{V}} \to \bar{\mathcal{V}} \bar{K}^{-2n} \, ,
\end{eqnarray}
\end{subequations}
then Eqs.~\eqref{eq:13} simply read
\begin{subequations}
\label{eq:21}
  \begin{eqnarray}
    \bar{\mathcal{M}^\prime} &=& x^{2} \left[ \theta^n \left( 1 + \frac{\theta^n}{2\bar{\lambda}} \right) + \frac{\bar{\mathcal{V}}}{\bar{\lambda}} \right] \ , \label{eq:21a} \\
    \theta^\prime &=& - \frac{3}{x^2} \left( \frac{x^3 \bar{p}_{eff} + \bar{\mathcal{M}}}{1 - 6 \bar{\mathcal{M}}/x} \right) \frac{\left( 1+ \theta \right)}{(n+1)} \, , \label{eq:21b} \\
    \bar{\mathcal{V}}^\prime &=& - \frac{12}{x^2} \left( \frac{x^3 \bar{p}_{eff} + \bar{\mathcal{M}}}{1 - 6 \bar{\mathcal{M}}/x} \right) \bar{\mathcal{V}} \nonumber \\
&& + \frac{9}{x^2} \left( \frac{x^3 \bar{p}_{eff} + \bar{\mathcal{M}}}{1 - 6 \bar{\mathcal{M}}/x} \right) \frac{ n ( 1 + \theta )^2}{(n+1) } \theta^{2n-1} \, , \label{eq:21c}
  \end{eqnarray}
\end{subequations}
where
\begin{equation}
\label{eq:22}
  \bar{p}_{eff} = \theta^{n+1} \left(1 + \frac{\theta^n}{\bar{\lambda}}  \right) + \frac{\theta^{2n}}{2\bar{\lambda}} + \frac{\bar{\mathcal{V}}}{3\bar{\lambda}} \, .
\end{equation}

Our main interest are the numerical solutions of stars with a finite
size, as determined by the boundary condition $p(X)=0$, which in the
case of the polytropes translates into $\rho(X) =0$, and from this
into $\theta(X)=0$. In order to avoid any singularities in the
equations of motion at the surface of the star, in particular for
Eq.~\eqref{eq:21c}, we must constraint the values of the polytropic
index in the range $n \geq 1/2$. Needless to say, such a constraint
does not exist either in the case of non-relativistic (Newtonian) or
relativistic (GR) polytropes. It must be noticed as well that the boundary conditions should also be adjusted so that $X = (G_NM/R)^{1/2} \bar{K}^{-n/2}$ and $\bar{\mathcal{M}}(X) = (1/3) (G_N M/R)^{3/2} \bar{K}^{-n/2}$.

We now include brane corrections with the contribution of one of the Weyl terms, with the boundary condition $\bar{\mathcal{V}}(X)=0$, so that the ID matching condition~\eqref{eq:9} allows a Schwarzschild exterior for the polytrope and dictates that the interior solution for the nonlocal energy density is given by Eq.~\eqref{eq:19}. As discussed in Sec.~\ref{sec:high-low-energy}, the brane terms must contribute to the effective density and pressure inside the star, which means that we cannot have in this case a counterpart of the GM solution, unless the Schwarzschild condition were waived.

As in the cases studied in Sec.~\ref{sec:case-const-dens}, we will
integrate inwards the equations of motion under the same boundary
conditions presented in Sec.~\ref{sec:boundary-conditions-}, with the central value of
$\theta(0)$ being a free parameter that will help us to classify the
numerical solutions. The most compact star will be given by the
maximum in the plot of the compactness as a function of the central
density: $G_N M/R$ vs $\theta(0)$. The value of the compactness will be read off from the outermost points of the numerical solution as $G_N M/R = 3 \bar{\mathcal{M}}(X)/X$, whereas the polytropic coefficient can be calculated from: $\bar{K} = [3 \bar{\mathcal{M}}(X)]^{1/3}/X$.

This time we have to give explicit values to the brane tension $\bar{\lambda}$ and the polytropic index $n$. For the latter, we consider in the following sections the case of white dwarfs with $n=3$, whereas the brane tension will remain free to label the different brane star solutions.

%%%%%%%%%%%%%%%%%%%%%%%%%%%%%%%%%%%%
\subsection{Numerical solutions \label{sec:numerical-solutions-}}
\begin{figure}[htp]
\includegraphics[width=0.5\textwidth]{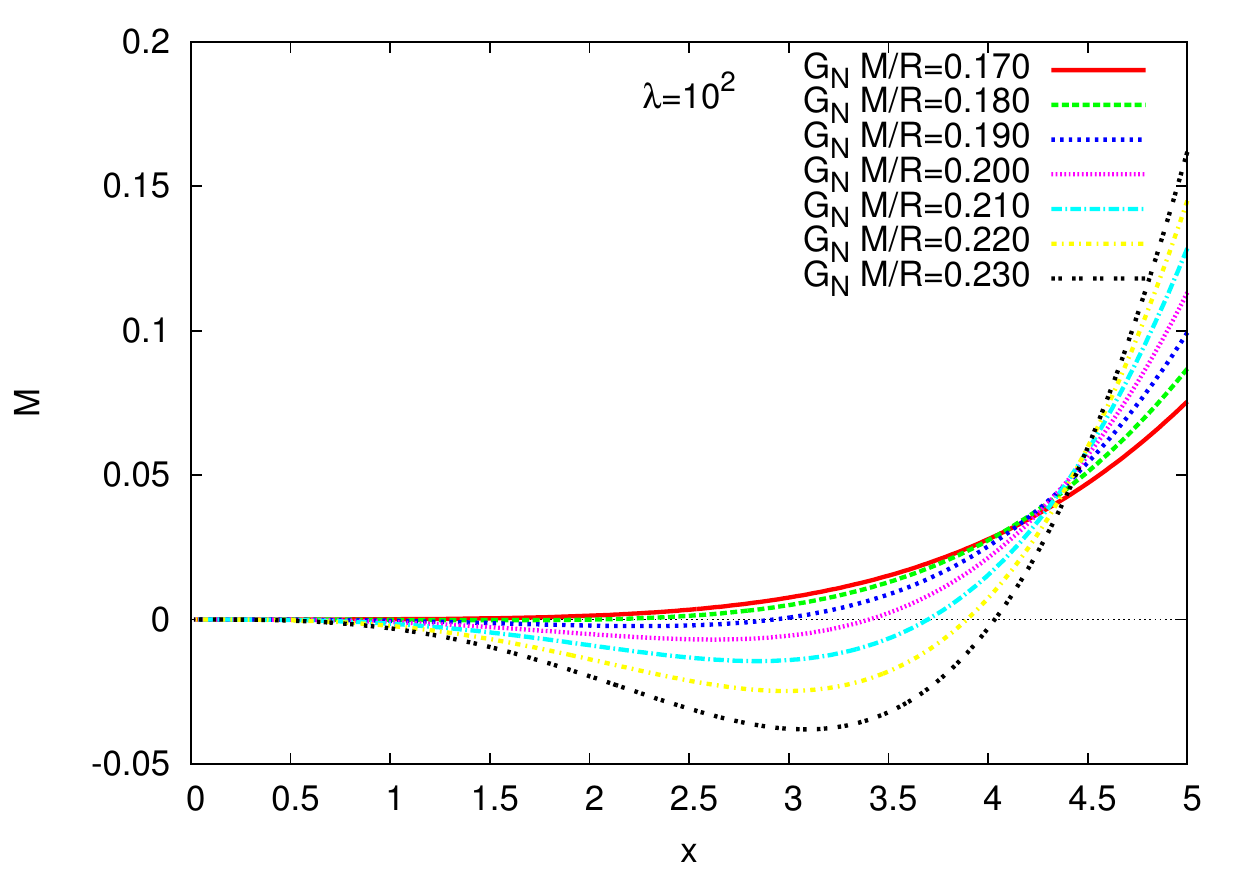}
\caption{\label{mass} Numerical solutions of the interior profile of
  the mass $M$ of polytropic brane stars, see Eqs.~\eqref{eq:21}, with
  $\lambda=10^2$. We can observe for $G_{N}M/R \geq 0.180$ that the
  mass becomes negative, like in the constant density case shown in
  Fig.~\ref{fig:GMWconstant}. This is due to the contribution of the
  Weyl function $\bar{\mathcal{V}}$, see text for more details.}
\end{figure}

\begin{figure}[htp!]
\includegraphics[width=0.5\textwidth]{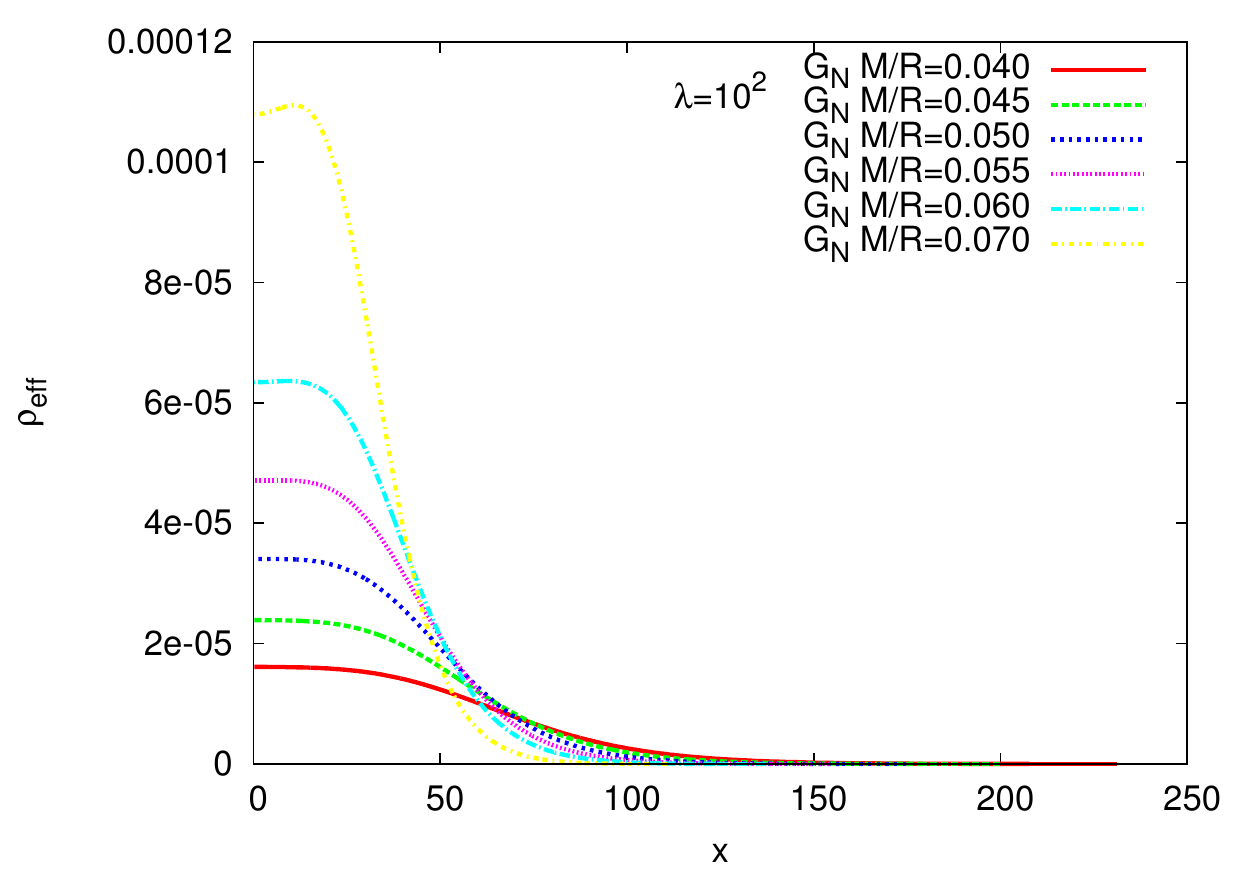} 
\includegraphics[width=0.5\textwidth]{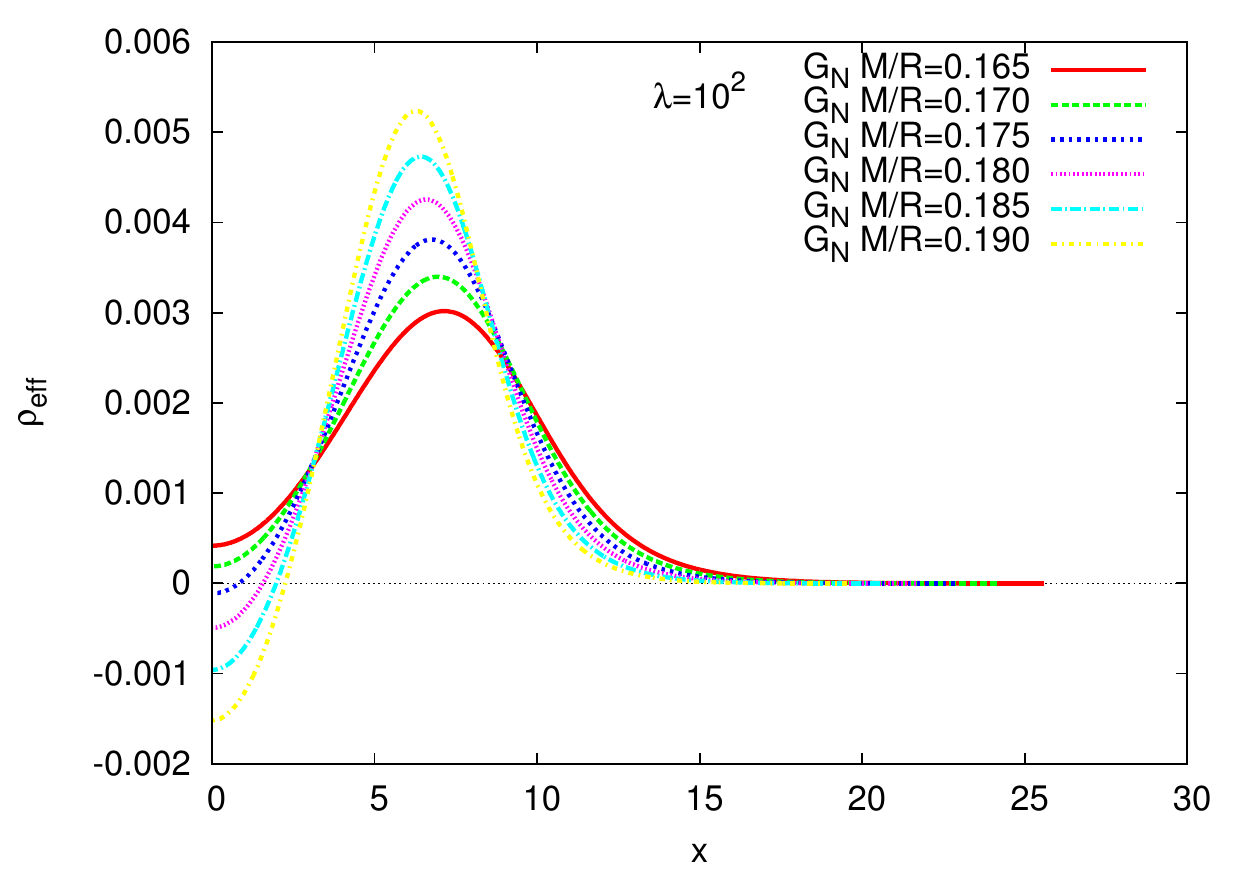}
\caption{\label{den} (Top) The effective energy density
  $\bar{\rho}_{eff}$ as a function of the radial coordinate $x$, which
  results from the numerical solution of Eqs.~\eqref{eq:21} for low
  compactness and $\lambda=10^2$. It can be seen that, as the
  compactness increases, the maximum of the effective density is
  displaced from the center of the star. (Bottom) The interior
  profiles of the effective density $\bar{\rho}_{eff}$ for high
  compactness. The effect of the Weyl term $\bar{\mathcal{V}}$ is
  sufficiently large to change the sign of the effective density at
  the center; actually, $\bar{\rho}_{eff}(x=0)< 0$ for $G_{N}M/R \geq
  0.175$.}
\end{figure}

%%%%%%%%%%%%%%%%%%%%%%%%%%%%%%%%%%%%
Solutions for the high energy limit with $\lambda=10^{2}$ allowed by
Eqs.~\eqref{eq:21} are shown in Figs.~\ref{mass} and~\ref{den}. In
particular, the interior mass profile $M(x)$ is shown in
Fig.~\ref{mass} for a range of compactness: $G_{N}M/R =0.170-0.230$,
in which the main feature we can observe is a change in sign close to
the center of the star for $G_{N}M/R \geq 0.180$. This behavior is
due to the contribution of the Weyl function $\bar{\mathcal{V}}$ in
Eq.~\eqref{eq:21a}. Note that the same behavior occurs in the constant
density case in Section~\ref{sec:extended-gm-solution-1}, so this type
of effect from the Weyl function is present also in more realistic
stars.

On the other hand, numerical solutions with $\lambda=10^2$ in
Fig.~\ref{den} show that for low compactness the effective density has
the expected decreasing behavior as we move outwards from the center
of the star. However, as the compactness increases the maximum value
of the density is displaced from the center, and then the density
profile is not just a decreasing function. Also note that the
effective density at the center becomes negative for $G_{N}M/R \geq
0.175$. It is clear that the geometric term $\bar{\mathcal{V}}$
contributes notoriously for large values of the compactness in the
high energy limit.

\begin{figure}[htp!]
\includegraphics[width=0.5\textwidth]{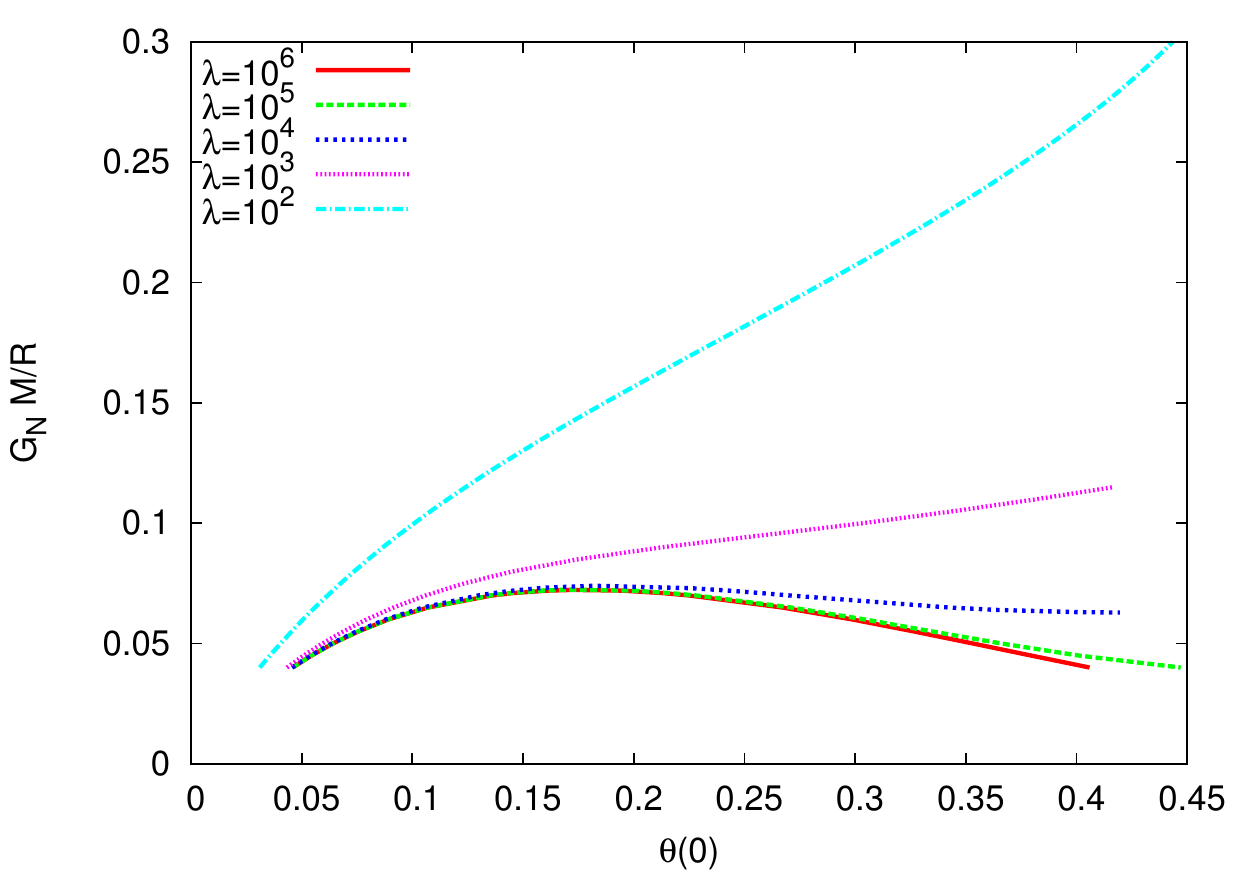}
\caption{\label{comp} The compactness $G_N M/R$ as a
  function of the central value $\theta(0)$ for the polytropic
  configurations obtained from Eqs.~\eqref{eq:21}. It can be seen that
  the compactness of the polytropic star with $n=3$ is not bounded as
  $\lambda \rightarrow 0$. For $\lambda=10^{6}, 10^5$ the curves coincides
  with that of polytropes in GR.}
\end{figure}

Thus, at least for the range of compactness we numerically explored,
the contribution of the Weyl tensor affects the internal
configurations of the stars in such a way that there is no maximum for
the compactness that can be reached, except for the Schwarzschild
bound $G_NM/R < 0.5$. This can be seen in Fig.~\ref{comp}, where we
note that for $\lambda=10^{6}, 10^{5}$ the curves reach a maximum
value just as in the case of polytropic stars in GR. However, as
$\lambda$ decreases it is possible to find stellar configurations with
a larger compactness beyond the standard GR bound.

\section{Conclusions and Remarks} \label{Disc}
%%%%%%%%%%%%%%%%%%%%%%%%%%%%%%%%%%%%%%%%%%%%%%%

In this paper we studied the equilibrium configurations of stars with
gravitational corrections in braneworld models, and provided numerical
solutions when necessary. For that we considered the high and low
energy limits of the equations of motion to show the threshold between
GR and braneworlds, and explored the appropriate boundary conditions
to obtain general conclusions about the physical properties of the
different stellar configurations. 

Our analysis took into account the corresponding Weyl functions which
provide non-local terms in the pressure and density, and which can
have noticeable effects in diverse features of a star. This study
allows us to relinquish the nonlocal anisotropic stress under the
conditions of a Schwarzschild exterior and non-constant density, which
are conditions rightly expected for a real star. 

As initial test, we revisited the case of constant density,
corresponding to the GM solution, but later studied the so-called
\emph{extended} GM solution, for which exist stars with a compactness
beyond the standard GR bound. This is due mainly to the existence of
the non-local terms which provoke the appearance of negative values of
the effective density and mass in the interior of the star.

Finally, we considered the case of a white dwarf star modeled with a
polytropic EoS and index $n=3$. In similarity with the extended GM
case, our results proved the existence of dwarf stars with a
compactness beyond the GR limit, because of the presence of non-local
terms. Also, the compactness of dwarf stars is not bounded as the
brane tension tends to zero, which correspond to the high energy
limit, while in the low energy limit we recovered the classical
compactness reported in the literature for the case of GR. All these
results are in agreement with the study in\cite{Garcia-Aspeitia:2014pna},
where it was shown that one of the main assumptions for the existence
of an upper bound in the compactness of a star was that the effective
energy density in the interior should be a decreasing function.

The results presented were based on a clear methodology that make the
equations of motion of braneworlds more tractable in numerical terms
that in other analyses in the literature. As noted, the presence of
brane corrections modifies in a notorious way the compactness, mass,
and other physical characteristics in stellar dynamics, even under the
assumption of a Schwarzschild exterior. Extended studies along the
lines suggested in this paper can be used to constrain the value of
the brane tension using observational data provided by stellar
dynamics, and with that to find evidence for the presence of extra
dimensions. As a final note, we cannot say if all the configurations
found would be gravitationally stable, but it is very likely that
those with a negative effective density in the interior may not be
able to prevent the collapse into configurations well within the
general bound found in\cite{Garcia-Aspeitia:2014pna}. This is work in
progress that will be reported elsewhere.

\begin{acknowledgements}
%%%%%%%%%%%%%%%%%%%%%%%%%%%%%%%%%%%%%%%%%%%%%%%
MAG-A acknowledges support from C\'atedra-CONACYT and SNI, also thanks
the Departamento de F\'isica-UG for its kind hospitality. This work
was partially supported by PROMEP, DAIP (534/2015), PIFI, and by
CONACyT M\'exico under grants  232893 (sabbatical), 167335 and 179881. We also thank the support of the Fundaci\'on Marcos Moshinsky, the Instituto Avanzado de Cosmolog\'ia (IAC), and the Beyond Standard Theory Group (BeST) collaborations.
\end{acknowledgements}

\bibliography{librero1}

\end{document}